\documentclass[12pt]{iopart}

\usepackage{graphicx}
\usepackage{bm}
\usepackage{cite}
\usepackage{color} 

\begin{document}

\title[Author guidelines for IOP journals in  \LaTeXe]{Influence of interface condition on spin-Seebeck effects}

\author{Z. Qiu$^1$, D. Hou$^1$, K. Uchida$^{2, 3}$,  and E. Saitoh$^{1, 2, 4, 5}$}

\address{$^1$ WPI Advanced Institute for Materials Research, Tohoku University, Sendai 980-8577, Japan}
\address{$^2$ Institute for Materials Research, Tohoku University, Sendai 980-8577, Japan}
\address{$^3$ PRESTO, Japan Science and Technology Agency, Saitama 332-0012, Japan}
\address{$^4$ CREST, Japan Science and Technology Agency, Tokyo 102-0076, Japan}
\address{$^5$ Advanced Science Research Center, Japan Atomic Energy Agency, Tokai 319-1195, Japan}

\ead{qiuzy@imr.tohoku.ac.jp}

\begin{abstract}
The longitudinal spin-Seebeck effect (LSSE) has been investigated for Pt/yttrium iron garnet (YIG) bilayer systems. The magnitude of the voltage induced by the LSSE is found to be sensitive to the Pt/YIG interface condition. We observed large LSSE voltage in a Pt/YIG system with a better crystalline interface, while the voltage decays steeply when an amorphous layer is introduced at the interface artificially. 
\end{abstract}

\pacs{85.75.−d, 72.25.−b, 72.15.Jf, 73.50.Lw} 
\noindent{\it Keywords}: Interface structure, spin-Seebeck effect, thermoelectric conversion, spin current
\maketitle

\section{Introduction}
The spin-Seebeck effect (SSE) enables the generation of a non-equilibrium spin current by a temperature gradient\cite{Slonczewski1996L1,  Adachi2010,  Uchida2011, Uchida2008, Jaworski2010, Bosu2011,  PhysRevLett.106.186601, Uchida2010a, Uchida2010, Kirihara2012, Kikkawa2013, Kikkawa2013a, Padron-Hernandez2011, Lu2012,DaSilva2013, Jungfleisch2013, Uchida2012, Cunha2013, Schreier2013,PhysRevB.83.094410,Xiao2010}. When a conductor is attached to a magnetic material, the SSE induces a spin-current injection into the conductor. The SSE is of crucial importance in spintronics \cite{Zutic2004,Wolf2001} and spin caloritronics \cite{PhysRevB.35.4959,Bauer2012}, since it enables simple and versatile generation of spin currents from heat. The spin current induced by the SSE is converted into electric voltage via the inverse spin-Hall effect (ISHE) owing to the spin orbit interaction in the conductor\cite{PhysRevB.83.094410,Xiao2010,Qiu2013a,Iguchi2013,Qiu2013,Qiu2012,Hou2012,Harii2012,Adachi2013,Saitoh2006,Kimura2007a,Valenzuela2006}. The combination of SSE and ISHE provides a new way of thermoelectric conversion. Although this phenomenon has been observed in many magnetic materials \cite{Uchida2008,Jaworski2010,Bosu2011,PhysRevLett.106.186601,Uchida2010a,Uchida2010,Kirihara2012,Kikkawa2013,Kikkawa2013a,Padron-Hernandez2011}, the thermoelectric conversion efficiency of the SSE device is still small and its improvement is essential for realizing future spin-current-driven thermoelectric generators\cite{Kirihara2012}.

 The thermoelectric voltage in a SSE device is affected by several factors, such as the spin Hall angle of the conductor, the quality and magnetic propeties of the magnetic material, and the conductor/magnet interface condition. The conductor/magnet interface condition in SSE devices should govern the injection efficiency of a thermal spin current across the interface, which is described in terms of the spin-mixing conductance\cite{Schreier2013,Xiao2010}. In our previous work, we found that the spin-mixing conductance of  a Pt/yttrium iron garnet (YIG) spin-pumping system is sensitive to the atomic-scale interface condition, and a well-controlled and highly-crystalline interface is a prerequisite for a large spin-mixing conductance\cite{Qiu2013}. However, few works have involved the influence of the interface condition on the SSE\cite{Miao2013}, and it is yet to be investigated systematically.
 
In this work, we systematically study the influence of the interface condition on the SSE in Pt/YIG systems, which have been widely used for investigating the SSE \cite{Uchida2010a, Uchida2010, Kirihara2012, Kikkawa2013, Kikkawa2013a,  Padron-Hernandez2011, Lu2012, DaSilva2013, Jungfleisch2013, Uchida2012, Cunha2013}. We controlled the Pt/YIG interface condition in atomic scale, and measured the ISHE voltage induced by the SSE with changing the interface conditions. These measurements enable to form a perceptual understanding of the influence of the interface condition on the SSE and of the spin-injection mechanism at the conductor/magnet interface.
 
 \section{Experimental}
 
\begin{figure}
\centering
\begin{minipage}[b]{1\textwidth}
\centering
\includegraphics[width=3in]{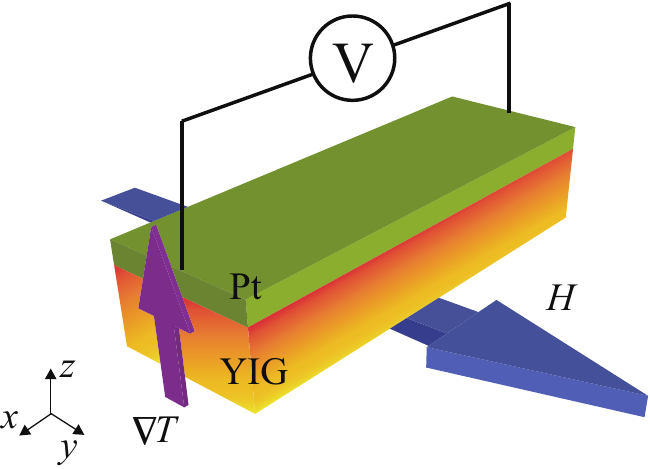}
\end{minipage}
\caption{A schematic illustration of LSSE. $\nabla T$ and $H$ denote the temperature gradient and the magnetic field, respectively.
\label{figure1}}
\end{figure}

 In this study, we employ a longitudinal configuration to investigate the SSE in the Pt/YIG system\cite{Uchida2010a,Uchida2012}. The longitudinal SSE (LSSE) device consists of a ferromagnetic or ferrimagnetic insulator covered with a paramagnetic conductor film (Fig. \ref{figure1}). When a temperature gradient $\nabla T$ is applied to the ferromagnetic layer perpendicular to the paramagnet/ferromagnet interface, a spin current is injected into the paramagnetic layer via the SSE. Then an ISHE-induced electric field $\bm{{\rm{E}}}_{\rm{ISHE}}$ is generated in the paramagnetic layer according to the relation
 \begin{equation}
\bm{{\rm{E}}}_{\rm{ISHE}} \propto\bm{{\rm{J}}}_{ \rm{s }}\times\bm{{\rm{\sigma}}} ,
\label{eq1}
\end{equation}
where $\bm{{\rm{J}}}_{ \rm{s }}$, and $\rm{\bm{\sigma}}$ denote the spatial direction of the thermally generated spin current and the spin-polarization vector of electrons in the paramagnetic layer, respectively. 

The preparation process of our Pt/YIG samples is as follows. The YIG films were grown on (111) gadolinium gallium garnet (GGG) substrates by using a liquid phase epitaxy (LPE) method in $\rm{PbO}$-$\rm{B_2O_3}$ flux at the temperature of 1210 K. The thickness of the YIG films is about 4.5 $\mu$m. The surfaces of the YIG films were annealed under oxygen pressure of $5\times 10^{-5}$ Torr at 1073 K for 2 hours to improve the crystal structure, or were bombarded by ion beams to create amorphous layers\cite{Qiu2013}. After the annealing or ion beam bombardment process, 10 nm-thick Pt films were deposited on those YIG films with the same condition by using a pulse laser deposition system at room temperature in the same vacuum chamber. The crystalline characterization of the Pt/YIG samples was carried out by means of an X-ray diffraction (XRD) method and a high-resolution transmission electron microscopy (TEM). The lattice constant was calculated from the XRD and electron diffraction patterns. The magnetization ($M$)-magnetic field ($H$) curve of the YIG film was measured by using a vibrating sample magnetometer. 

The LSSE measurements were performed by using an experimental system similar to that described in Ref. \cite{Uchida2012}. The Pt/YIG sample with the size of $2\times 6$ mm$^2$ was sandwiched between two AlN heat baths, of which the temperatures were stabilized to 300 K+$\Delta T$ and 300 K, respectively. The temperature gradient was generated by a Peltier thermoelectric module. The electric voltage difference $V$ between the ends of the Pt layer was measured by using a micro-probing system, where a distance between the two probes is about 5 mm.

\section{Results and discussion}

\begin{figure}
\centering
\begin{minipage}[b]{1\textwidth}
\centering
\includegraphics[width=6in]{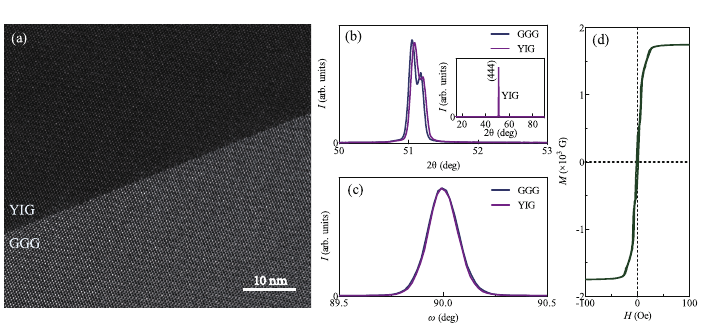}
\end{minipage}
\caption{(a) A cross-sectional high-resolution TEM image of the YIG/GGG interface. (b) The comparison of the XRD patterns for the YIG film and the GGG substrate. The inset shows the XRD pattern for the YIG film in a large $2\theta$ region. (c) The comparison of the rocking curves between the YIG film and the GGG substrate. (d) The $M$-$H$ curve for the YIG film.
\label{figure2}}
\end{figure}

First of all, we evaluate the quality of the YIG films grown by the LPE method. Figure \ref{figure2}(a) shows the TEM image of the YIG/GGG interface; the image confirms that the YIG film is epitaxially grown on the GGG substrate, and no defect is observed owing to the small lattice-constant mismatch between YIG and GGG\cite{Geller1957,Carruthers1973}. The XRD patterns of the YIG film in Fig. \ref{figure2}(b) show a good single crystal characteristic, where only the (444) peak is observed. The lattice constants of the YIG film and GGG substrate determined by the XRD results are 12.376 $\rm{\mathring {A}} $ and 12.384  $\rm{\mathring {A}}$, respectively, both of which are in good agreement with the values found in previous literature\cite{Geller1957,Carruthers1973}. The YIG films have almost the same crystal quality as the GGG substrate because of the same full width at half maximum of the rocking curves (Fig. \ref{figure2}(c)). The $M$-$H$ curve of the YIG film in Fig. \ref{figure2}(d) shows that the coercive force of the YIG film is very small (smaller than 5 Oe) and the saturation magnetization is 4$\pi M_s$=1710 G (close to the bulk value \cite{Shone1985}). These sample characterizations clearly confirm that our YIG films are of very high quality.

\begin{figure}
\centering
\begin{minipage}[b]{1\textwidth}
\centering
\includegraphics[width=6in]{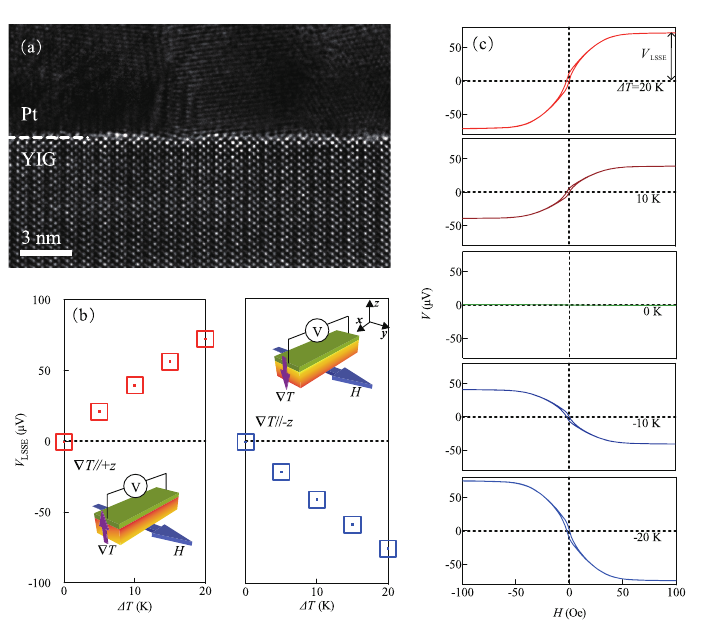}
\end{minipage}
\caption{(a) A cross-sectional high-resolution TEM image of the Pt/YIG interface, of which the YIG was annealed before the Pt layer was formed. (b) The $\Delta T$ dependence of $V_{\rm{LSSE}}$ for the Pt/YIG LSSE device, shown in (a), measured when a temperature gradient $\nabla T$ is applied along the $+z$ (left) and $-z$ (right) directions. (c) $H$ dependence of $V$ for the Pt/YIG LSSE device, shown in (a), for various values of $\Delta T$. 
\label{figure3}} 
\end{figure}

Figure \ref{figure3}(a) shows the cross-sectional structure of the prepared Pt/YIG interface measured by a high-resolution TEM. The crystal perfection of the YIG surface was found to be well kept even after forming the Pt film; the YIG layer has a perfect single-crystalline structure with a [111] direction perpendicular to the Pt/YIG interface. The Pt layer has a typical polycrystalline structure without a preferred orientation. The Pt/YIG interface used in this study has the same high quality as that shown in Ref. \cite{Qiu2013}. By using such Pt/YIG devices, we measured the LSSE voltage $V_{\rm{LSSE}}$. 

In Fig. \ref{figure3}(b), we show $V_{\rm{LSSE}}$ in the Pt/YIG sample with the highly-crystalline interface as a function of the temperature difference $\Delta T$ at the magnetic field $H$=100 Oe, measured when the magnetic field was applied along the $+y$ direction. As shown in  Fig. \ref{figure3}(b), the sign of $V_{\rm{LSSE}}$ for finite values of  $\Delta T$ in the Pt/YIG device is reversed by reversing the $\nabla T$ direction. This result indicates that the observed voltage signals are attributed to the LSSE, on the basis that the direction of the thermally generated spin current at the Pt/YIG interface is reversed by reversing $\Delta T$. Figure  \ref{figure3}(c) shows the voltage signal $V$ as a function of $H$ for various values of $\Delta T$. The sign of $V$ is reversed by reversing $H$ at finite values of $\Delta T$, consistent with the feature of the ISHE-induced electric field (see Eq. (\ref{eq1})). The shape of the $V$-$H$ curves is in good agreement with that of the $M$-$H$ curve for the YIG film (compare Figs. \ref{figure2}(b) and \ref{figure3}(c)), confirming that the observed voltage $V$ is associated with the magnetization reversal of the YIG layer. 

\begin{figure}
\centering
\begin{minipage}[b]{1\textwidth}
\centering
\includegraphics[width=6in]{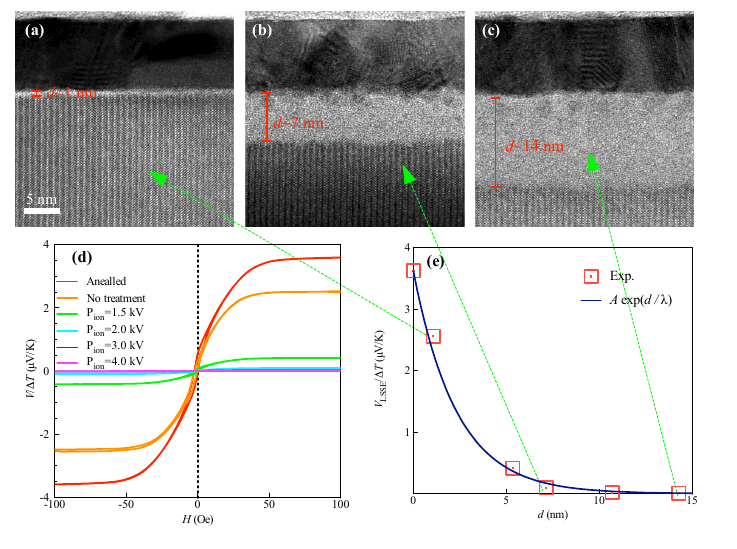}
\end{minipage}
\caption{(a)-(c) The cross-sectional high-resolution TEM images of the Pt/YIG samples, of which the YIG films were prepared without any anealling process (a),  with the ion beam bombardment at 2 kV-acceleration voltage (b), and with the ion beam bombardment at 4 kV-acceleration voltage (c). The ion bombardment was performed before the Pt layers were formed. (d) The $H$ dependence of $V/\Delta T$ in the Pt/YIG devices with various interfaces. (e) The dependence of $V_{\rm{LSSE}}/\Delta T$ on the amorphous-layer thickness $d$. The solid curve is the fitting result using an exponential function $A \rm{exp}$ $(d/\lambda)$, where $A$ and $\lambda$ are adjustable parameters. 
\label{figure4}}
\end{figure}

Now we investigate the influence of the Pt/YIG interface condition on the LSSE voltage. To compare the magnitude of the LSSE voltage, we normalize $V_{\rm{LSSE}}$ with the temperature difference  $\Delta T$. The $V_{\rm{LSSE}}/\Delta T$ for the Pt/YIG sample with the highly-crystalline interface is about 3.6 $\rm{\mu V/K}$. This value is 45\% larger than that for the sample without the annealing process (Fig. \ref{figure4}(d)), in which an amorphous layer with the thickness of $<$1 nm was observed at the Pt/YIG interface (Fig. \ref{figure4}(a)). This enhancement of $V_{\rm{LSSE}}/\Delta T$ is attributed to the improvement of the spin mixing conductance, which determines the transfer efficiency of spin currents across the Pt/YIG interface. \cite{Qiu2013}

To further investigate the relation between the interface condition and the LSSE voltage, we measure the dependence of $V_{\rm{LSSE}}$ on the thickness of the amorphous interlayer in the Pt/YIG samples. We found that the thickness of the amorphous layer at the Pt/YIG interface can be changed by ion-beam bombardment; the thickness of the amorphous layer increases with increasing the acceleration voltage ($\rm{P_{ion}}$) of the ion beam (Fig. \ref{figure4} (a), (b) and (c)). Although it is difficult to quantitatively evaluate the properties of the amorphous layers, we believe that its macroscopic magnetization is extremely depressed because of the broken crystal structure, while localized spins may be alive because of the presence of iron atoms.

The $H$ dependence of $V_{\rm{LSSE}}/\Delta T$ for various amorphous-layer thicknesses are plotted in Fig. \ref{figure4}(d).  The magnitude of $V_{\rm{LSSE}}/\Delta T$ in the Pt/YIG samples decreases exponentially with increasing the thickness $d$ of the amorphous layer, indicating that the amorphous layer blocks the spin injection from YIG to Pt. By fitting the $d$ dependence of $V_{\rm{LSSE}}/\Delta T$ with an exponential function $A \rm{exp} $ $(d/\lambda)$ with $A$ being an adjustable parameter, the decay length $\lambda$ is estimated to be $\sim $2.3 nm. The estimated $\lambda$ value for the amorphous layer is greater than those for other non-magnetic insulator oxide barriers estimated from the spin-pumping measurements using Pt/oxide/YIG systems\cite{Du2013}, implying that the remaining localized spins in the amorphous layer may reduce a potential barrier for spin currents\cite{Ren2014}. Nevertheless, our experimental results show that clean Pt/YIG interfaces are necessary for realizing efficient spin injection.

\section{Conclusion}
In this work, the influence of the interface crystal structure on the SSE has been investigated by using Pt/YIG bilayer LSSE devices. We found that the ISHE voltage induced by the LSSE is very sensitive to the interface condition of the YIG film. The large LSSE voltage was found to be achieved by forming a highly-crystalline Pt/YIG interface in an atomic scale. Our systematic experiments show that the magnitude of the LSSE voltage decreases exponentially with increasing the thickness of the artificial amorphous layer formed at the top surface of the YIG film.
  
\section*{Acknowledgements}


This work was supported by CREST ``Creation of Nanosystems with Novel Functions through Process Integration'', PRESTO ``Phase Interfaces for Highly Efficient Energy Utilization'', Strategic International Cooperative Program ASPIMATT from JST, Japan, Grant-in-Aid for Young Scientists (B) (26790038), Young Scientists (A) (25707029), Challenging Exploratory Research (26600067), Scientific Research (A) (24244051), Scientific Research on Innovative Areas ``Nano Spin Conversion Science'' (26103005) from MEXT, Japan, the Tanikawa Fund Promotion of Thermal Technology, the Casio Science Promotion Foundation, and the Iwatani Naoji Foundation.

\section*{References}


\end{document}